\documentclass[11pt]{article}
\usepackage{amsmath}
\usepackage{amssymb}
\usepackage{amsthm}
\usepackage{amsfonts}
\usepackage{comment}
\usepackage{color}
\usepackage{mathrsfs}
\usepackage{braket}
\usepackage{graphicx}
\usepackage[subrefformat=parens]{subcaption}
\usepackage{cite}
\usepackage{here}
\usepackage{tikz}
\usetikzlibrary{decorations.pathmorphing}
\usepackage{caption}

\usepackage[stable]{footmisc}

\makeatletter

\@addtoreset{equation}{section}
\makeatother

\begin{document}

\pagestyle{empty}

\begin{flushright}
\end{flushright}

\vspace{3cm}

\begin{center}

{\bf\LARGE  
Metric fluctuations\\in higher-dimensional black holes
}

\vspace*{1.5cm}
{\large 
Hyewon Han\footnote{dwhw101@dgu.ac.kr}, Bogeun Gwak\footnote{rasenis@dgu.ac.kr}
} \\
\vspace*{0.5cm}

{\it 
Division of Physics and Semiconductor Science, Dongguk University, Seoul 04620,\\Republic of Korea
}

\end{center}

\vspace*{1.0cm}

\begin{abstract}
{\noindent
We investigated the impact of metric fluctuations on the higher-dimensional black hole geometry. We generalized the four-dimensional model to higher dimensions to treat quantum vacuum fluctuations by the classical approach. A fluctuating black hole is portrayed by a higher-dimensional Vaidya metric with a spherically oscillating mass. Assuming a small fluctuation amplitude, we employed a perturbation method to obtain a radially outgoing null geodesic equation up to the second order in the fluctuation. Furthermore, the fluctuation of the event horizon up to the second order depends on the number of spacetime dimensions. Therefore, the time-averaged values of the thermodynamic variables defined at the horizon also feature dimension-dependent correction terms. A general solution was obtained for rays propagating near the horizon within a fluctuating geometry. Upon examining this in a large $D$ limit, we found that a complete solution can be obtained in a compact form.}

\end{abstract}

\newpage
\baselineskip=18pt
\setcounter{page}{2}
\pagestyle{plain}
\baselineskip=18pt
\pagestyle{plain}

\setcounter{footnote}{0}
	
\section{Introduction}
Einstein's classical theory of gravity  posits that a black hole, resulting from gravitational collapse and settling into a stationary state, is characterized by three asymptotically observable parameters: mass, charge, and angular momentum. Contrary to the classical notion of a black hole solely absorbing matter and energy, quantum mechanical effects in proximity to the black hole horizon trigger a thermal energy transition from the black hole over long distances. Quantum fluctuations of the black hole metric, spurred by the uncertainty principle, initiate the spontaneous creation of energy quanta and the steady emission of thermal energy to infinity, giving the black hole the semblance of a body with temperature $\kappa / 2 \pi $ (in Planck units), where $\kappa$ is the black hole's surface gravity \cite{hawking1975particle,bekenstein1973black}. The intersection of the fields of gravity, quantum mechanics, and thermodynamics in the context of black hole thermodynamics has been an active research area in recent decades.

Dynamic descriptions of this Hawking effect were provided by York \cite{york1983dynamical}, who classically approximated the zero-point fluctuations that induce the black hole's thermal quantum radiation to an ingoing Vaidya-type metric with a mass oscillating in the black hole's quasinormal mode. He showed that such quantum fluctuations lead to the formation of a `quantum ergosphere,’ and calculated the black hole entropy and thermal fluctuations via a statistical method of the quasinormal modes.
The statistical origins of the black hole entropy and its thermodynamic properties have been the subject of extensive investigation \cite{york1986black,frolov1993dynamical,demers1995black,ho1997entropy,strominger1998black,wu2004entropy,ghosh2011black,wang2018entropy,arzano2019quantum,sinha2022hawking}.
Moreover, the quasinormal modes of black holes play a significant role in understanding their dynamics and quantum properties \cite{cardoso2001quasinormal,berti2003asymptotic,berti2003highly,zerbini2004asymptotics,andersson2004asymptotic,berti2004black,cho2006asymptotic,dolan2009expansion,corda2015time,jaramillo2021pseudospectrum,Yang:2021civ,Gwak:2022mze,mamani2022revisiting,kyutoku2023quasinormal}.

The effects of black hole metric fluctuations on Hawking radiation have been studied by Barrab$\grave{\mathrm{e}}$s et al.\cite{barrabes1999metric}. Employing the York model, they solved the equation for outgoing null rays perturbed by spherically symmetric fluctuations, deriving corrections for the outgoing energy flux and asymptotic spectrum of the s-waves. Fluctuating geometry modifications can also be found in \cite{bellucci2010thermodynamic,arias2012thermal,frolov2017quantum}.
Additionally, the fluctuations of other quantum fields interacting with the gravitational field were approximated using a stochastic ensemble of metric fluctuations. When such `induced' fluctuations dominate, stochastic gravity theory can address the fluctuations and back-reaction problems of dynamical black hole spacetime \cite{barrabes2000stochastically,hu2007metric,hu2008stochastic}.

In the pursuit of a complete description of quantum effects in the gravitational field, spacetime of more than four dimensions has garnered considerable attention \cite{emparan2008black,horowitz2012black}. The concept of extra dimensions was introduced for mathematical consistency in string theory which serves as one of the promising candidates for quantum gravity. The brane-world scenario, assumed to resolve the problem of gravity’s scale being much larger than the scale of the electroweak force, posits that all standard model matters are confined to a four-dimensional hypersurface embedded in higher-dimensional spacetime, while gravity propagates through large extra dimensions \cite{kanti2004black}. The AdS/CFT correspondence links $D$-dimensional quantum field theory to a $(D+1)$-dimensional gravity theory \cite{aharony2000large}. The diverse dynamics in higher-dimensional spacetime offer valuable insights for a more comprehensive understanding of quantum gravity.

Interestingly, when the number of spacetime dimensions is large, the complex nonlinear dynamics of gravity theory are significantly simplified \cite{emparan2013large,emparan2015effective,emparan2020large}. The gravitational influence of a black hole is sharply localized in a thin region near the horizon, allowing for independent analysis from the region far from the black hole. As a result, the dynamics of the black hole horizon can be reduced to a simple and effective theory. Recent research utilizing this ‘large $D$ effective theory’ has examined various black hole models \cite{licht2020black,suzuki2021squashed,suzuki2022rotating}, and investigated the phases and instabilities of black strings/branes \cite{li2021ads,licht2022lattice,licht2022large,suzuki2023phase}. The large $D$ limit offers a rich landscape for analysis, furnishing new perspectives \cite{andrade2020entropy,mandlik2021black,mandlik2022sitter,kachru2023holographic,keeler2022hidden,giataganas2022holographic,kirezli2022classification,sybesma2023zoo,luna2023holographic}.

In this work, we investigated the influence of metric fluctuations in higher-dimensional black hole geometry. By integrating the effects of black hole radiation, it is possible to provide a more consistent quantum gravitational description of geometry. We generalize the four-dimensional models of York \cite{york1983dynamical} and Barrab$\grave{\mathrm{e}}$s \cite{barrabes1999metric} to higher dimensions, creating a simple model for treating fluctuations by the classical approach. This model features an asymptotically flat, static higher-dimensional black hole metric with a spherically oscillating source. Assuming the amplitude of the fluctuations to be minuscule for a massive black hole, we analyze this geometry using classical perturbation theory. By solving the perturbed outgoing null geodesic equation in arbitrary $D$-dimensions, we investigated the dimensional dependency of the position of the black hole horizon and thermodynamic quantities. We then calculated the general corrections for the perturbed rays propagating near the horizon. Ultimately, we demonstrate that the perturbation terms are significantly simplified for a large $D$, enabling compact, complete solutions to be obtained.

The remainder of this paper is organized as follows. 
In Section 2, we provide a brief review of York's and Barrab$\grave{\mathrm{e}}$s models, as well as higher-dimensional Vaidya black holes.
In Section 3, we generalize the model to higher dimensions and calculate the outgoing null geodesic equation.
In Section 4, the positions of the perturbed horizon and time-averaged thermodynamic variables are obtained.
In Section 5, we examine the propagation of outgoing rays in a fluctuating geometry.
In Section 6, we apply the large $D$ limit.
Section 7 offers a summary of our study's findings. 
Throughout this study, we employ a metric signature of $(-, +, +, +)$ and use dimensionless units \cite{barrabes1999metric} such that $c=G=\hbar=1$.

\section{Review on model and the higher-dimensional Vaidya black hole}
\noindent
The model proposed by York \cite{york1983dynamical} represents a four-dimensional black hole undergoing simple zero-point oscillations near its horizon. This model is approximated using an ingoing Vaidya-type metric.
        \begin{align}
            ds^2 = - \left( 1- \frac{2m}{r} \right) d v^2 + 2 dv dr + r^2 d\theta^2 + r^2 \sin^2\theta d\phi^2,
        \end{align}
where
        \begin{align} \label{eq:mass}
            m = m(v,\theta)=M_{BH} \left[ 1 + \sum_l (2l+1) \mu_l h_l(v)q_l(\theta) \right].
        \end{align}
The function $m(v,\theta)$ serves as the source of the fluctuating geometry, $M_{BH}$ represents the total mass of the black hole, $l$ is the azimuthal quantum number, $q_l$ is a spherical harmonic with zero magnetic quantum number (to be normalized), and $\mu_l$ is a dimensionless amplitude parameter of the oscillations. The function $h_l(v)$ and the amplitude parameter are assumed to have the form
        \begin{align}
            h_l(v)= \sin \omega_l v, \qquad \mu_l = \alpha_l \frac{M_{Planck}}{M_{BH}},
        \end{align}
where $\omega_l$ is the oscillation frequency, and $\alpha_l$ is a pure number. Note that $\mu_l \ll 1$ for a black hole with mass $M_{BH} \gg M_{Planck}$. This metric describes a Vaidya black hole fluctuating at a small amplitude.

Barrab$\grave{\mathrm{e}}$s \cite{barrabes1999metric, Barrabes:2013zva} studied the effects of a fluctuating geometry on outgoing rays using this classical treatment of fluctuations. They considered a simple case of spherical oscillations with $l=0$, yielding $q_0(\theta)=1$ and assumed that the black hole was formed from the gravitational collapse of a spherical massive null shell. Hence, the mass function can be written as
        \begin{align}
            m = m (v) = M_{BH} \left[1+\mu(v) \right] \theta (v) = M_{BH} \left[1+\mu_0 \sin ( \omega v) \right] \vartheta (v),
        \end{align}
where $\vartheta (v)$ is the Heaviside step function, which indicates that a spherical null shell with mass equal to $M_{BH}$ collapses at the origin of the advanced time coordinate $v=0$. In this geometry, radial null rays depart from past null infinity $\cal{J}^-$ at an advanced time $v$ and arrive at future null infinity $\cal{J}^+$ at a retarded time $u$. For large $u$, outgoing rays propagate near the event horizon and their effective frequencies are very high compared to the characteristic frequency of the black hole, $\kappa = 1/4M$. This allows us to use geometric optics approximation to study the propagation of null rays near the horizon. For $v<0$, rays leaving $\cal{J}^-$ propagate inward in flat spacetime, bounce off at $r=0$ (the regular center of the flat space), and propagate outward. After crossing the null shell at $v=0$, the rays propagate in the black hole spacetime and reach $\cal{J}^+$ at large $u$. Figure \ref{fig:diagram} schematically shows the conformal structure and propagation of radial null rays in the absence of fluctuations.

\begin{center}
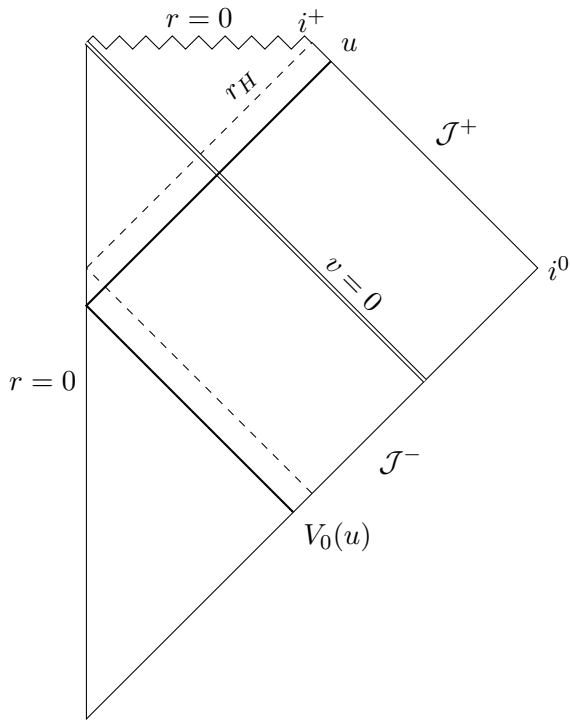

\begin{tikzpicture} 

    \node (I)    at (3,0)   {};
    \node (a)    at (3,-1.5) {};

    \path
     (I) +(90:3)  coordinate[label=90:$i^+$] (Itop)
         +(-90:3) coordinate (Ibot)
         +(180:3) coordinate (Ileft)
         +(0:3)   coordinate[label=0:$i^0$] (Iright)
       ;
    \draw[dashed] (Ileft) -- node[near end, above, sloped] {$r_H$}
                  (Itop);
    \draw (Itop) --
             node[midway, above right]    {$\cal{J}^+$}      
          (Iright) -- 
             node[near end, below right]    {$\cal{J}^-$}
          (Ibot);
    \draw[dashed]
          (Ibot) --  (Ileft);
      
    \path 
     (Itop) + (-3,0) coordinate (sing);
    \path 
      (sing) +(0,-9) coordinate (past);
    \draw (sing) -- node[midway, left]    {$r=0$}
          (past) -- (Ibot) ;
    \draw[decorate,decoration=zigzag] (sing) -- (Itop)
         node[midway, above, inner sep=2mm] {$r=0$};

    \path
      (a) +(1.5,0) coordinate (v0);
    \draw[thin,double distance=1pt]
        (v0) -- 
          node[near start, above, sloped] {$v=0$}
        (sing);

    \path
      (Ileft) +(0,-0.5) coordinate (bounce);
    \path
     (Ibot)  +(-0.25,-0.25) coordinate[label=-85:$V_0(u)$] (Vu);
    \path
     (Itop) +(0.25,-0.25) coordinate[label=20:$u$] (rad);
    \draw[thick]
     (Vu) -- (bounce) -- (rad);
    
\end{tikzpicture} 
\captionof{figure}{Conformal diagram of the spherically symmetric black hole formed by a gravitational collapse of a null shell. $\cal{J}^-$ denotes past null infinity, $\cal{J}^+$ denotes future null infinity, and the points $i^+$ and $i^0$ denote future timelike infinity and spatial infinity, respectively. The double line $v=0$ represents the collapsing null shell and the dark solid line represents the trajectory of the null ray.}
\label{fig:diagram}
\end{center}

\noindent
By tracing the trajectory of the ray reaching $\cal{J}^+$ at $u$ backward in time, we obtain the advanced time $v=V_0(u)$ when the ray starts at $\cal{J}^-$. In the presence of metric fluctuations, the causal structure of this geometry remains unchanged. However, $v=V_0(u)$ is modified to a new value, $v=V(u)$, because the trajectory of the ray is modified. Therefore, the relation $V(u)$ provides information on the fluctuating geometry and is used to study the modified energy flux and asymptotic spectrum of radiation.

In this study, we are interested in the effects of black hole metric fluctuations in higher dimensions. For a simple model, we considered a non-rotating, neutral, and higher-dimensional black hole. The Schwarzschild-Tangherlini metric\cite{tangherlini1963schwarzschild} is an asymptotically flat and static black hole solution derived from higher-dimensional Einstein's theory of gravity in vacuum. This metric is given by \cite{emparan2008black}
        \begin{align}
        \label{eq:STmetric}
            ds^2 = - \left( 1- \frac{M}{r^{D-3}} \right) d t^2 + \frac{ d r^2}{\left( 1- \frac{M}{r^{D-3}} \right)} + r^2 d \Omega_{D-2}^2,
        \end{align}
where $d \Omega_{D-2}^2$ is the line elements on a ($D-2$)-dimensional unit sphere $S^{D-2}$. Here $M$ is a mass parameter which has following relation with the black hole mass $M_{BH}$. 
        \begin{align} \label{eq:massparameter}
            M_{BH} = \frac{(D-2)\Omega_{D-2}}{16 \pi} M, 
        \end{align}
where $\Omega_{D-2}=\frac{2\pi^{(D-1)/2}}{\Gamma(\frac{D-1}{2})}$ represents the volume of $S^{D-2}$. We assume that the source of black holes fluctuates over time. The Vaidya-type metric is useful in describing massive spherical bodies with oscillating or radiating null fluids. We use ingoing Eddington-Finkelstein coordinates $(v,r)$, where $v = t + r_{*}$ is the advanced time, allowing the mass parameter in \eqref{eq:STmetric} to become a function of $v$. Hence, the higher-dimensional Vaidya metric \cite{iyer1989vaidya} is written as
        \begin{align}
        \label{eq:hVaidya}
            ds^2 = - \left( 1- \frac{m(v)}{r^{D-3}} \right) d v^2 + 2 dv dr + r^2 d \Omega_{D-2}^2.
        \end{align}
The radial tortoise coordinate $r_*$ is defined for this metric as
        \begin{align}
        \label{eq:tortoise}
            \frac{d r_{*}}{d r} = \left( 1- \frac{m(v)}{r^{D-3}} \right)^{-1}.
        \end{align}
The event horizon of the black hole is located at
        \begin{align}
            r_H = \left[ \frac{m(v)}{1-2(d r_H / d v)}\right]^{\frac{1}{D-3}}.
        \end{align} 
The surface area of the horizon and the surface gravity are given by
        \begin{align}
            \mathcal{A} = \Omega_{D-2} r_H^{D-2}(v), \qquad \kappa =  \frac{D-3}{2 r_H (v)}.
        \end{align}
Using these quantities and thermodynamic relations, we can derive the Hawking temperature $T_H = \kappa / 2\pi$ and the entropy $S = \mathcal{A} / 4$ of the black hole. We apply the mass function $m(v)$ following York and Barrab$\grave{\mathrm{e}}$s to classically treat the metric fluctuations and study how these variables and the trajectory of the radial null rays are modified in this geometry.

\section{The fluctuating geometry in higher dimensions}
\noindent
In this section, we construct a simple model through generalization to higher dimensions. We employ the metric \eqref{eq:hVaidya} to consider a higher $D$-dimensional spacetime and utilize the fluctuating mass function to observe the effects of small perturbations on the metric. Because the function \eqref{eq:mass} contains a three-dimensional spherical harmonic $q_l(\theta)$, we must generalize it to higher dimensions.
        \begin{align} 
            m = m(v,\theta) = M \left[ 1 + \sum_l (2l+1) \mu_l \sin (\omega_l v) Y_l(\theta) \right] \vartheta (v).
        \end{align}
where $M$ is the higher-dimensional mass parameter defined by \eqref{eq:massparameter} and $Y_l(\theta)$ is the hyperspherical harmonic (which is assumed to have an appropriate normalizing constant applied) that has zero magnetic quantum number. Note that $\vartheta (v)$ indicates that the black hole was formed by the collapse of a null shell of mass $M$. In general, the hyperspherical harmonics for $D>3$ are given by \cite{zhao2017spherical, erdelyi1953higher}
        \begin{align}
            Y_l ( m_k;\theta, \phi) = e^{\pm \mathrm{i} m_{D-3} \phi} \prod^{D-4}_{k=0} \left( \sin \theta_{D-k-3} \right)^{m_{k+1}} \mathrm{C}^{m_{k+1}+\frac{D-k-3}{2}}_{m_k-m_{k+1}} \left( \cos \theta_{D-k-3} \right)^{m_{k+1}},
        \end{align}
where $m_k$ is the magnetic quantum number and
        \begin{align}
            \mathrm{C}^a_b (t) = \frac{(2a)_b}{b!}  \left. _2F_1 \left(-b, b+2a, a+\frac{1}{2}; \frac{1-t}{2} \right) \right.
        \end{align}
is the Gegenbauer polynomials. For the simplest case with no dependence of $\theta$ we consider the lowest mode $l=m_k=0$. In that case, 
        \begin{align}
            Y_0 = \prod^{D-4}_{k=0} \frac{(D-k-3)_0}{0!}  \left. _2F_1 \left(0, D-k-3, \frac{D-k-2}{2}; \frac{1-\cos \theta_{D-k-3}}{2} \right) \right.=1.
        \end{align}
Consequently, we obtain the following fluctuating mass function.
        \begin{align}
            m(v) = M \left[1+\mu(v) \right] \vartheta (v) = M \left[1+\mu_0 \sin ( \omega v) \right] \vartheta (v) ,
        \end{align}
where $\mu_0$ is a dimensionless small-amplitude parameter and $\omega$ is the frequency of fluctuations (the subscript 0 is dropped). The second term in parentheses is associated with fluctuations. By substituting the metric \eqref{eq:hVaidya} with this mass function into Einstein's equation, the energy-momentum tensor of this geometry is obtained as
        \begin{align}
            T_{a b} = \frac{(D-2)}{16 \pi r^{D-2}} \left[ M (1+\mu_0 \sin ( \omega v)) \delta(v) + M\mu_0 \omega \cos(\omega v) \vartheta(v) \right] l_a l_b,
        \end{align}
where $l_a = - \partial_a v$ denotes the null vector field tangent to the radially ingoing null geodesics.

To study the effect of metric fluctuations on the propagation of null rays, we solve the equation for radial rays. Setting $d s^2 = 0$ and $d \Omega_{D-2} = 0$ gives 
        \begin{align}
            - \left( 1- \frac{m(v)}{r^{D-3} } \right) d v^2 + 2 \, d v d r = 0.
        \end{align}
The ingoing radial rays move with $d v = 0$ and the outgoing radial rays move with $d u = 0$, or
        \begin{align} \label{eq:ingoing}
            \left( 1- \frac{m(v)}{r^{D-3} } \right) d v = 2 \, d r.
        \end{align}
Because the dimensionless amplitude $\mu_0$ of the fluctuations is assumed to be very small, we use the perturbation method.
        \begin{align} \label{eq:per}
            r = r (v) = R (v) + \rho (v) + \sigma (v) + \cdots
        \end{align}
where $R(v)$ is the unperturbed solution or the solution in the absence of metric fluctuations, and $\rho(v)$ and $\sigma(v)$ are the first- and second-order perturbations of $\mu_0$, respectively. We ignore the higher orders in $\mu_0$ denoted by dots here. 
We investigated the modified trajectory of the ray in a fluctuating geometry by fixing $u$ as a constant value of the retarded time when the ray reached $\cal{J}^+$ in the absence of fluctuations.
        \begin{align}
            u = t - R_*(v) = v - 2R_*(v) = constant. 
        \end{align}
Here, $R_*$ is an unperturbed radial tortoise coordinate defined by
        \begin{align}
            \frac{d R_{*}}{d R} = \left( 1- \frac{M}{R^{D-3}} \right)^{-1},
        \end{align}
which is given by \cite{berti2004gravitational}
        \begin{align}
                R_* = R - M^{\frac{1}{D-3}} + \frac{M}{D-3} \sum^{D-4}_{j=0} \frac{\ln\left(R/\alpha_j-1\right)}{\alpha_j^{D-4}},
        \end{align}
and 
        \begin{align}
            \alpha_j=M^{\frac{1}{D-3}} e^{\mathrm{i} \frac{2 \pi}{D-3} j}.
        \end{align}
        
When $v<0$, the outgoing ray propagates in flat spacetime with $d v = 2 d R$. For $v>0$, substituting the perturbed radial coordinates \eqref{eq:per} into equation \eqref{eq:ingoing} and linearizing, we obtain 
        \begin{align}
            2 \, \frac{d R }{d v} &= 1 - \frac{M}{R^{D-3} } , \label{eq:zeroth}\\
            2 \, \frac{d \rho }{d v}& - (D-3) \frac{M}{R^{D-2} } \rho = - \frac{M}{R^{D-3} } \mu , \label{eq:first1} \\ 
            2 \, \frac{d \sigma }{d v}& - (D-3) \frac{M}{R^{D-2} } \sigma = (D-3) \frac{M}{R^{D-3} } \left[ \frac{\rho \mu}{R} - \frac{D-2}{2} \frac{\rho^2}{R^2} \right]. \label{eq:second1}
        \end{align}
We can write the equations for the perturbations \eqref{eq:first1}--\eqref{eq:second1} as
        \begin{align} \label{eq:perturbationeq}
            \frac{d f}{d v} = \frac{D-3}{2} \frac{M}{R^{D-2} } f + F.
        \end{align}
The first-order perturbation corresponds to
        \begin{align} \label{eq:first2}
            f = \rho \, \,,\, F = - \frac{1}{2} \frac{M}{R^{D-3} }  \mu,
        \end{align}
and the second-order perturbation corresponds to
        \begin{align} \label{eq:second2}
            f = \sigma \, \,,\, F = \frac{D-3}{2} \frac{M}{R^{D-3} } \left[ \frac{\rho \mu}{R} - \frac{D-2}{2} \frac{\rho^2}{R^2} \right].
        \end{align}
These equations include the results in \cite{barrabes1999metric} for $D=4$. We solve these perturbed null-ray equations to obtain a solution that describes a fluctuating event horizon of the black hole and the general trajectory of the ray propagating near the horizon.

\section{The perturbed event horizon and thermodynamics}
We seek a particular solution describing the position of the perturbed horizon $r_H (v) = R_H (v) + \rho_H (v) + \sigma_H (v) + \cdots$ and examine the quantities defined on the horizon in the presence of fluctuations. From the zeroth-order equation \eqref{eq:zeroth}, the unperturbed horizon is located at
        \begin{align}
            R_H = M^{\frac{1}{D-3}}.
        \end{align}
Following this, the unperturbed surface gravity of the black hole is 
        \begin{align} \label{eq:surfacegravity}
            \kappa =  \frac{D-3}{2 R_H} = \frac{D-3}{2 M^{\frac{1}{D-3}}}.
        \end{align}
By using these quantities and equations \eqref{eq:first2}--\eqref{eq:second2}, we find that
        \begin{align}
            \frac{d \rho_H }{d v} - \kappa \rho_H &= - \frac{1}{2} \mu, \\
            \frac{d \sigma_H }{d v} - \kappa \sigma_H &= \kappa \rho_H \mu- \frac{D-2}{2} \frac{\kappa \rho^2_H}{R_H}.
        \end{align}       
Upon solving these equations, we obtain the position of the horizon in the fluctuating geometry, 
        \begin{align}
            \rho_H = & \frac{\mu_0}{2 \kappa} \frac{ \Omega \cos \omega v + \sin \omega v}{1+ \Omega^2} ,\\
            \nonumber \\ 
            \sigma_H =& \frac{\mu_0^2}{4 \kappa} \left[ \frac{ 2\Omega^2 (2 - \Omega^2) \cos 2\omega v + \Omega(1-5\Omega^2) \sin 2 \omega v}{(1+ \Omega^2)^2 (1+4\Omega^2)} \right. \nonumber \\
            & \qquad \qquad \left. - \frac{D-4}{D-3} \frac{ (1-5\Omega^2) \sin^2 \omega v + \Omega (2-\Omega^2) \sin 2 \omega v +   \Omega^2(5+2\Omega^2) }{(1+ \Omega^2)^2 (1+4\Omega^2)} \right],
        \end{align}
where $\Omega=\omega/\kappa$ denotes a dimensionless frequency. The integral constant was chosen to eliminate terms causing small perturbations to grow exponentially over time. The position of the horizon $r_H = M^{\frac{1}{D-3}} + \rho_H + \sigma_H$ changes periodically, depending on the sine and cosine functions, with frequencies and amplitudes determined by the metric fluctuation parameters. Thus, one can imagine the horizon wriggling due to fluctuations. The second term in the second-order correction $\sigma_H$ depends on the number of dimensions $D$ and vanishes for four dimensions.

Next, we calculate the mean values of the fluctuating surface area, surface gravity, Hawking temperature, and entropy of the black hole, averaged over time $v$ to determine the overall change. The average values of the surface area and surface gravity on the fluctuating horizon are
        \begin{align}
            \overline{\mathcal{A}} \equiv \Omega_{D-2}\, \overline{ r_H^{D-2}(v)\, } = \Omega_{D-2} \, R_H^{D-2} \left[1+\frac{\mu_0^2 (D-2)}{4 (1+\Omega^2)(D-3)^2} \right], \\
            \nonumber \\
            \overline{\kappa} \equiv \frac{D-3}{2} \overline{\left(\frac{m(v)}{r_H^{D-2}(v)}\right)}  = \kappa \left[1+ \frac{\mu_0^2 (D-2)}{4 (1+\Omega^2)(D-3)^2}\right],
        \end{align}
where $\kappa$ is the value of the unperturbed surface gravity \eqref{eq:surfacegravity}. In the presence of metric fluctuations, these quantities have slightly larger values than their counterparts in the absence of fluctuations. Furthermore, the higher the dimensions $D$, the smaller the second term's value, corresponding to the effect of the fluctuations. The Hawking temperature was modified to 
        \begin{align}
            \overline{T}_H=\frac{\overline{\kappa}}{2 \pi} = \frac{\kappa}{2 \pi} \left[1+\frac{\mu_0^2 (D-2)}{4 (1+\Omega^2)(D-3)^2}\right].
        \end{align}
The changes in surface area $\delta \mathcal{A} = \overline{\mathcal{A}} - \mathcal{A}$ and Hawking temperature $\delta T_H = \overline{T}_H - T_H$ due to fluctuations satisfy the following relation
        \begin{align}
            \frac{\delta \mathcal{A}}{\mathcal{A}} = \frac{\delta T_H}{T_H}.
        \end{align}
This relationship holds not only in four dimensions but also in higher dimensions. The standard relation $S = \mathcal{A}/4$ for the entropy of a black hole is also modified. The average entropy value was obtained using the first law $d \overline{S} = d E / \overline{T_H}$, where $E=\overline{m}(v)=M$.
        \begin{align}
            \overline{S} &=  \int \frac{2 \pi}{\kappa} d E \left(1+ \frac{\mu_0^2 (D-2)}{4 (1+\Omega^2)(D-3)^2}\right)^{-1} \simeq \frac{\overline{\mathcal{A}}}{4} \left(1- \frac{\mu_0^2 (D-2)}{2 (1+\Omega^2)(D-3)^2}\right).
        \end{align}
Again, the impact of the fluctuations diminishes as the dimension $D$ increases.

\section{Propagation of perturbed radial rays}
\noindent
We herein solve the null geodesic equations for the general case $R \ne R_H$ after obtaining radially outgoing null geodesic equations for a fluctuating higher-dimensional black hole geometry and examining how the event horizon fluctuates. For $v<0$, the outgoing ray propagates with $d u = 0$ in flat geometry. For $v>0$, the trajectory of the outgoing ray is given by equations \eqref{eq:zeroth}--\eqref{eq:second1}. Using equation \eqref{eq:zeroth}, we change the variable $v$ in equation \eqref{eq:perturbationeq} to the unperturbed trajectory $R(v)$, yielding the solution
        \begin{align}
                f = \left(1-\frac{M}{R^{D-3}}\right) \left[ - \int^{\infty}_{R} \frac{2F}{\left(1-\frac{M}{R'^{D-3}}\right)^2} d R' + f_0 \right],
        \end{align}
where $f_0$ is the integration constant. We assume that the outgoing ray reaches future null infinity $\cal{J}^+$ at the same time $u$ in both the absence and presence of fluctuations. Therefore, the perturbation term $f$ should vanish at $\cal{J}^+$. For both $\rho$ and $\sigma$, $F$ approaches zero as $R\to \infty$, and we satisfy this requirement by setting $f_0=0$.
We introduce the following dimensionless quantities for convenience.
        \begin{align}
                x = \frac{R-R_H}{R_H}, \qquad \tilde{u}=\kappa u, \qquad \tilde{f} = \frac{2}{d-3} \kappa f.
        \end{align}
Subsequently, we express the dimensionless perturbations as
        \begin{align} \label{eq:gp1}
            \tilde{\rho}(x) = \left[1-\frac{1}{(1+x)^{D-3}}\right] \int^{\infty}_{x} \frac{\mu(\xi) \, (1+\xi)^{D-3}}{\left\{(1+\xi)^{D-3}-1\right\}^2} d\xi,
        \end{align}
        \begin{align} \label{eq:gp2}
            \tilde{\sigma}(x) = - \left[1-\frac{1}{(1+x)^{D-3}}\right] \int^{\infty}_{x} \frac{(1+\xi)^{D-3}}{\left\{(1+\xi)^{D-3}-1\right\}^2} (D-3) \left\{ \frac{\tilde{\rho}(\xi)\mu(\xi)}{1+\xi} - \frac{D-2}{2} \frac{\tilde{\rho}(\xi)^2}{(1+\xi)^2} \right\} d\xi.
        \end{align}
We also express the fluctuating mass term in the integrand in dimensionless form as
        \begin{align} \label{eq:fmass}
                \mu (\xi) &= \mu_0 \sin (\omega \, v(\xi)) = \mu_0 \sin \left[ \Omega (\tilde{u} + 2 \kappa R_*(\xi) \right] \nonumber \\
                &=\mu_0 \sin \left[ \Omega \left( \tilde{u} + (D-3)\xi + \sum^{D-4}_{j=0} e^{\mathrm{i} \frac{2\pi}{D-3}j}\ln\left(\frac{1+\xi}{e^{\mathrm{i} \frac{2\pi}{D-3}j}}-1\right)\right)\right].
        \end{align}

These solutions describe the modified trajectory of the outgoing radial null ray in a higher-dimensional fluctuating black hole geometry. Now, we establish the relationship $v=V(u)$ between the null coordinates to observe the entire history of the perturbed rays in this geometry. We first consider a situation with no fluctuations ($\mu_0=0$). For $v<0$, the spacetime is flat, and the null coordinates are related by $v-u=2R$. The rays leaving $\cal{J}^-$ at $v=V_0<0$ converge toward the regular origin $R=0$ of spacetime, bounce off at $R=0$, and propagate outward. Because the ingoing rays move with $v=constant$ and the outgoing rays move with $u=constant$, we obtain the relation 
        \begin{align} \label{eq:V01}
            V_0=-2R_0,
        \end{align}
where $R_0$ denotes the values of the unperturbed radial coordinates when $v=0$. For $v>0$, the null coordinates are related by $v-u=2R_*$ in the black hole spacetime. In this region, the rays propagate outward only with $u=constant$, and we obtain 
        \begin{align} \label{eq:V02}
            -u = 2R_{*0} = 2\left[R_0 - M^{\frac{1}{D-3}} + \frac{M^{\frac{1}{D-3}}}{D-3} \sum^{D-4}_{j=0} e^{\mathrm{i} \frac{2\pi}{D-3}j} \ln\left(\frac{R_0}{M^{\frac{1}{D-3}}e^{\mathrm{i} \frac{2\pi}{D-3}j}} -1\right)\right].
        \end{align}
The null shell propagating along $v=0$ is a singular null hypersurface across which the metric is continuous. Thus, combining equation \eqref{eq:V01} obtained in the $v<0$ region with equation \eqref{eq:V02} obtained in the $v>0$ region yields the relationship $V_0(u)$.

We now consider the rays in the presence of fluctuations ($\mu_0 \ne 0$): Because we fixed the value of the retarded time $u$ at which the unperturbed rays reach $\cal{J}^+$, equation \eqref{eq:V02} remains unchanged, but equation \eqref{eq:V01} should be modified to
        \begin{align}
                V(u) = - 2 \left[R_0(u) + \rho_0(u) + \sigma_0(u) \right].
        \end{align}
The subscript $0$ represents the intersection of the radial coordinates of the ray and null shell. In the dimensionless form, it can be written as 
        \begin{align} \label{eq:Vfunc}
            \tilde{V}(\tilde{u}) = \kappa V = - \left[(D-3)(1+x_0) + \tilde{\rho}(x_0) + \tilde{\sigma}(x_0) \right],
        \end{align}
where $x_0=\frac{R_0-R_H}{R_H}$ is the value of the dimensionless variable $x$ on null shell $v=0$. This implies that $x_0$ satisfies 
        \begin{align}
                \tilde{u} + (D-3)x_0 + \sum^{D-4}_{j=0} e^{\mathrm{i} \frac{2\pi}{d-3}j}\ln\left(\frac{1+x_0}{e^{\mathrm{i} \frac{2\pi}{d-3}j}}-1\right)=0,
        \end{align}
which yields $x_0(\tilde{u})$. By computing the values of the fluctuations $\tilde{\rho}(x_0)$ and $\tilde{\sigma}(x_0)$, we obtain the relationship $\tilde{V}(\tilde{u})$. This function illustrates the impact of metric fluctuations in higher $D$-dimensions.

\section{The fluctuating geometry in the large $D$ limit}
\noindent
Calculating the integrals in equations \eqref{eq:gp1} and \eqref{eq:gp2} for generally higher dimensions can be complex and difficult. Hence, exploring the boundaries of the dimension number $D$, a well-defined natural parameter in gravity theory, is a good strategy for simplifying the problem.
 This section discusses how the equations become straightforward in the `large $D$ limit' \cite{emparan2013large, emparan2020large} and provides the corrected relation $\tilde{V}(\tilde{u})$ in the fluctuating geometry.

In the large $D$($\gg 1$) dimension, the radial gradient of the gravitational potential at the black hole horizon becomes very large $(\sim D/R_H)$, thereby localizing the gravitational field of the black hole strongly near the horizon. Outside the thin area on the order of $1/D$ where the influence of gravity exists, the black hole geometry becomes a flat Minkowski spacetime. One can define the `near-horizon zone' to study an effective theory in the vicinity of the horizon. Since we are interested in the rays arriving at $\cal{J}^+$ at a late time $u\gg 1$, which propagate close to the horizon, we introduce near-horizon coordinates
        \begin{align}
            \hat{\mathsf{R}} \equiv \left(\frac{R}{R_H}\right)^{D-3},
        \end{align}
defined by $\ln \hat{\mathsf{R}} \ll D-3$, to compute the perturbed trajectory of a ray in a large $D$-dimension. Considering the first order in $D^{-1}$, we obtain the radial tortoise coordinate from
        \begin{align}
            d R_* = \frac{1}{1-\left(\frac{R_H}{R}\right)^{D-3}}\, d R = \frac{1}{1- \frac{1}{\hat{\mathsf{R}}}} \frac{R_H}{(D-3)} \frac{1}{\hat{\mathsf{R}}} \,d \hat{\mathsf{R}} = \frac{R_H}{(D-3)} \frac{1}{\hat{\mathsf{R}}-1} \, d \hat{\mathsf{R}},
        \end{align}
yielding
        \begin{align}
            R_* = \frac{R_H}{D-3} \ln (\hat{\mathsf{R}}-1).
        \end{align}
Subsequently, the fluctuating mass term \eqref{eq:fmass} becomes 
        \begin{align}
            \mu (\hat{\mathsf{R}}) = \mu_0 \sin \left[ \Omega \left( \tilde{u} + 2 \kappa R_* \right ) \right] = \mu_0 \sin \left[ \Omega \left( \tilde{u} +\ln (\hat{\mathsf{R}}-1) \right) \right] = \mu_0 \mathrm{I m} \left[ e^{i \Omega \tilde{u}}(\hat{\mathsf{R}}-1)^{i \Omega} \right] .
        \end{align} 
With near-horizon coordinates, the first-order perturbation in $\mu_0$ is written as
        \begin{align}
            \tilde{\rho} (\hat{\mathsf{R}}) = 2 \kappa \rho (\hat{\mathsf{R}})= \left(1-\frac{1}{\hat{\mathsf{R}}}\right) \, I(\hat{\mathsf{R}}),
        \end{align}
where
        \begin{align}
            I(\hat{\mathsf{R}}) &= \int^{\infty}_{\hat{\mathsf{R}}} \frac{\mu(\tau)}{(\tau - 1)^2}\,d \tau =  \mu_0 \mathrm{I m} \left[e^{i \Omega \tilde{u}} \int^{\infty}_{\hat{\mathsf{R}}} (\tau - 1)^{i \Omega -2}\,d \tau \right] \\ \nonumber
            &= \mu_0 \mathrm{I m} \left[ \frac{1 + i\Omega}{1+\Omega^2} e^{i \Omega \tilde{u}} (\hat{\mathsf{R}}-1)^{i \Omega -1}\right]
        \end{align}
can be easily integrated. We require the value $\tilde{\rho}_0$ on the null shell to compute the function $\tilde{V}(\tilde{u})$. Here, $v=0$ implies
        \begin{align} \label{eq:onnullshell}
            \tilde{u} + \ln (\hat{\mathsf{R}}_0-1) = 0,
        \end{align}
with $\hat{\mathsf{R}}_0$ denoting the value of $\hat{\mathsf{R}}$ on the null shell. Using this relationship, we obtain the first-order perturbation on the null shell as 
        \begin{align}
            \tilde{\rho}_0(\hat{\mathsf{R}}_0) &= \left(1-\frac{1}{\hat{\mathsf{R}}_0}\right) \mu_0 \mathrm{I m} \left[\frac{1 + i\Omega}{1+\Omega^2} (\hat{\mathsf{R}}_0-1)^{-i \Omega}(\hat{\mathsf{R}}_0-1)^{i \Omega -1}\right] \nonumber \\
            &= \mu_0 \frac{\Omega}{1+\Omega^2} \frac{1}{\hat{\mathsf{R}}_0}.
        \end{align}

Next, we express the second-order perturbation in $\mu_0$ in near-horizon coordinates as
        \begin{align}
            \tilde{\sigma} (\hat{\mathsf{R}}) = 2 \kappa \sigma (\hat{\mathsf{R}}) = - \left(1-\frac{1}{\hat{\mathsf{R}}}\right) \int^{\infty}_{\hat{\mathsf{R}}} \frac{1}{(\tau - 1)^2} \tilde{\rho}(\tau) \left\{ \mu(\tau) - \frac{1}{2} \tilde{\rho}(\tau)\right\}\,d \tau.
        \end{align}
By substituting $\tilde{\rho} (\tau) = \left(1-\frac{1}{\tau}\right) \, I(\tau)$, we obtain
        \begin{align}
            \tilde{\sigma} (\hat{\mathsf{R}}) = - \left(1-\frac{1}{\hat{\mathsf{R}}}\right) \left[\int^{\infty}_{\hat{\mathsf{R}}} \frac{\mu (\tau)}{\tau(\tau-1)} I(\tau) d\tau - \frac{1}{2} \int^{\infty}_{\hat{\mathsf{R}}} \frac{1}{\tau^2} I^2(\tau) d\tau \right].
        \end{align}
By integrating the second term by parts, we obtain  
        \begin{align}
            \int^{\infty}_{\hat{\mathsf{R}}} \frac{1}{\tau^2} I^2(\tau) d\tau = \frac{I^2(\hat{\mathsf{R}})}{\hat{\mathsf{R}}} + 2 \int^{\infty}_{\hat{\mathsf{R}}} \frac{I(\tau)}{\tau} I'(\tau) \,d\tau,
        \end{align}
where the prime symbol denotes the derivative with respect to the integration variable $\tau$. Then we have
        \begin{align}
            \tilde{\sigma} (\hat{\mathsf{R}}) = \left(1-\frac{1}{\hat{\mathsf{R}}}\right) \left[\frac{I^2(\hat{\mathsf{R}})}{2\hat{\mathsf{R}}} - \int^{\infty}_{\hat{\mathsf{R}}} \frac{\mu(\tau)}{(\tau-1)^2} I(\tau) \,d\tau \right].
        \end{align}
Using
        \begin{align}
            \int^{\infty}_{\hat{\mathsf{R}}} \frac{\mu(\tau)}{(\tau-1)^2} I(\tau) \,d\tau = \frac{1}{2} I^2(\hat{\mathsf{R}}),
        \end{align}
we obtain the second order correction
        \begin{align}
            \tilde{\sigma} (\hat{\mathsf{R}}) = - \frac{1}{2} \left[ \left(1-\frac{1}{\hat{\mathsf{R}}}\right) I(\hat{\mathsf{R}})\right]^2.
        \end{align}
Again, we require the value on the shell. Using condition \eqref{eq:onnullshell}, we calculate the function $I^2(\hat{\mathsf{R}}_0)$ as 
        \begin{align}
            I^2(\hat{\mathsf{R}}_0) = \mu_0^2 \frac{\Omega^2}{(1+\Omega^2)^2}\frac{1}{(\hat{\mathsf{R}}_0-1)^2}.
        \end{align}
Therefore, the second-order perturbation on the null shell is 
        \begin{align}
            \tilde{\sigma}_0 (\hat{\mathsf{R}}_0) = - \frac{\mu_0^2}{2}\frac{\Omega^2}{(1+\Omega^2)^2} \frac{1}{\hat{\mathsf{R}}_0^2}.
        \end{align}

Finally, we compute the corrected relation $\tilde{V}(\tilde{u})$ in a large $D$ dimension. By fixing the value of $\tilde{u}$ as the (dimensionless) retarded time parameter when the unperturbed ray reaches $\cal{J}^+$, we obtain
        \begin{align}
            \hat{\mathsf{R}}_0 = 1+e^{-\tilde{u}},
        \end{align}
from equation \eqref{eq:onnullshell}. In the presence of fluctuations, the relationship between the null coordinates of the ray is 
        \begin{align}
            - \tilde{V}(\tilde{u}) &= (D-3) + \ln \hat{\mathsf{R}}_0(\tilde{u}) + \tilde{\rho}_0(\hat{\mathsf{R}}_0(\tilde{u}))+\tilde{\sigma}_0(\hat{\mathsf{R}}_0(\tilde{u})) \nonumber \\
            & = \tilde{V}_0 + \mu_0 \frac{\Omega}{1+\Omega^2} (1+e^{-\tilde{u}})^{-1} - \frac{\mu_0^2}{2}\frac{\Omega^2}{(1+\Omega^2)^2} (1+e^{-\tilde{u}})^{-2},
        \end{align}
where
        \begin{align}
            \tilde{V}_0 = (D-3)+\ln\left(1+e^{-\tilde{u}}\right)
        \end{align}
is the value in the absence of fluctuations. The corrected terms contain the amplitude and frequency parameters determined by the parameters of the metric fluctuations. Interestingly, when the dimension $D$ is very large, a complete solution is obtained in a compact form in the near-horizon zone.

\section{Conclusions}
\noindent
In this work, we examined the impact of metric fluctuations on outgoing radial rays propagating near the horizon of a higher-dimensional black hole. By generalizing the works of York\cite{york1983dynamical} and Barrab$\grave{\mathrm{e}}$s et al.\cite{barrabes1999metric} to higher dimensions, we studied an ingoing Vaidya-type metric with a spherically oscillating source over time $v$. The oscillating mass induced fluctuations in the geometry, necessitating the correction of the null geodesic equation. Given that the oscillations in our model were minuscule compared to the black hole mass, we used a perturbation method to obtain the corrected equation of radially outgoing rays. The perturbed event horizon solution was found up to the second order in the small-amplitude parameter $\mu_0$, and the second-order correction exhibited a dimensional dependency arising from the higher-dimensional spacetime. The perturbed horizon influenced the thermodynamic variables defined on the horizon. By calculating the time averages, we examined the global changes in the presence of fluctuations. The corrections in these variables decreased as the number of spacetime dimensions increased, and a dimensional dependency emerged in the relationship between the entropy and surface area of the black hole. We also derived a general solution for the perturbed outgoing rays. 
It is worth noting that our results encompass those of \cite{barrabes1999metric} for four dimensions.
In the large $D$ limit, we were able to analytically solve the complex integrals present in the general solution. Using these results, we established a complete $V(u)$ function, representing the corrected relationship between the null coordinates of the perturbed ray in a large $D$-dimension.
Although our results were obtained using a simplified model that assumed classical perturbations, they offer intriguing insights into higher dimensions. 

By adjusting the $V(u)$ function, we can derive corrections for the outgoing energy flux and the asymptotic spectrum of Hawking radiation in higher dimensions. Moreover, our model can be generalized to the case of a higher-dimensional rotating black hole, and non-spherical oscillations can be considered.

\vspace{10pt} 

\noindent{\bf Acknowledgments}

\noindent This research was supported by Basic Science Research Program through the National Research Foundation of Korea (NRF) funded by the Ministry of Education (NRF-2022R1I1A2063176) and the Dongguk University Research Fund of 2023. BG appreciates APCTP for its hospitality during the topical research program, {\it Multi-Messenger Astrophysics and Gravitation}.\\

\bibliographystyle{jhep}
\bibliography{ref}

\providecommand{\href}[2]{#2}\begingroup\raggedright\begin{thebibliography}{10}

\bibitem{hawking1975particle}
S.~W. Hawking, \emph{Particle creation by black holes}, {\emph{Communications
  in mathematical physics} {\bfseries 43} (1975) 199}.

\bibitem{bekenstein1973black}
J.~D. Bekenstein, \emph{Black holes and entropy}, {\emph{Physical Review D}
  {\bfseries 7} (1973) 2333}.

\bibitem{york1983dynamical}
J.~W. York~Jr, \emph{Dynamical origin of black-hole radiance}, {\emph{Physical
  Review D} {\bfseries 28} (1983) 2929}.

\bibitem{york1986black}
J.~W. York~Jr, \emph{Black-hole thermodynamics and the euclidean einstein
  action}, {\emph{Physical Review D} {\bfseries 33} (1986) 2092}.

\bibitem{frolov1993dynamical}
V.~Frolov and I.~Novikov, \emph{Dynamical origin of the entropy of a black
  hole}, {\emph{Physical Review D} {\bfseries 48} (1993) 4545}.

\bibitem{demers1995black}
J.-G. Demers, R.~Lafrance and R.~C. Myers, \emph{Black hole entropy without
  brick walls}, {\emph{Physical Review D} {\bfseries 52} (1995) 2245}.

\bibitem{ho1997entropy}
J.~Ho, W.~T. Kim, Y.-J. Park and H.~Shin, \emph{Entropy in the kerr-newman
  black hole}, {\emph{Classical and Quantum Gravity} {\bfseries 14} (1997)
  2617}.

\bibitem{strominger1998black}
A.~Strominger, \emph{Black hole entropy from near-horizon microstates},
  {\emph{Journal of High Energy Physics} {\bfseries 1998} (1998) 009}.

\bibitem{wu2004entropy}
Y.-J. Wu, Z.~Zhao and X.-J. Yang, \emph{Entropy of a radiating rotating charged
  black hole}, {\emph{Classical and Quantum Gravity} {\bfseries 21} (2004)
  2595}.

\bibitem{ghosh2011black}
A.~Ghosh and A.~Perez, \emph{Black hole entropy and isolated horizons
  thermodynamics}, {\emph{Physical review letters} {\bfseries 107} (2011)
  241301}.

\bibitem{wang2018entropy}
X.-Y. Wang, J.~Jiang and W.-B. Liu, \emph{Entropy in the interior of a kerr
  black hole}, {\emph{Classical and Quantum Gravity} {\bfseries 35} (2018)
  215002}.

\bibitem{arzano2019quantum}
M.~Arzano, L.~Brocki, J.~Kowalski-Glikman, M.~Letizia and J.~Unger,
  \emph{Quantum ergosphere and brick wall entropy}, {\emph{Physics Letters B}
  {\bfseries 797} (2019) 134887}.

\bibitem{sinha2022hawking}
A.~K. Sinha, \emph{Hawking decay and thermodynamic transformation of a black
  hole: two examples}, {\emph{Journal of the Korean Physical Society}
  {\bfseries 80} (2022) 359}.

\bibitem{cardoso2001quasinormal}
V.~Cardoso and J.~P. Lemos, \emph{Quasinormal modes of schwarzschild--anti-de
  sitter black holes: Electromagnetic and gravitational perturbations},
  {\emph{Physical Review D} {\bfseries 64} (2001) 084017}.

\bibitem{berti2003asymptotic}
E.~Berti and K.~D. Kokkotas, \emph{Asymptotic quasinormal modes of
  reissner-nordstr{\"o}m and kerr black holes}, {\emph{Physical Review D}
  {\bfseries 68} (2003) 044027}.

\bibitem{berti2003highly}
E.~Berti, V.~Cardoso, K.~D. Kokkotas and H.~Onozawa, \emph{Highly damped
  quasinormal modes of kerr black holes}, {\emph{Physical Review D} {\bfseries
  68} (2003) 124018}.

\bibitem{zerbini2004asymptotics}
S.~Zerbini and L.~Vanzo, \emph{Asymptotics of quasinormal modes for
  multihorizon black holes}, {\emph{Physical Review D} {\bfseries 70} (2004)
  044030}.

\bibitem{andersson2004asymptotic}
N.~Andersson and C.~Howls, \emph{The asymptotic quasinormal mode spectrum of
  non-rotating black holes}, {\emph{Classical and Quantum Gravity} {\bfseries
  21} (2004) 1623}.

\bibitem{berti2004black}
E.~Berti, \emph{Black hole quasinormal modes: hints of quantum gravity?},
  {\emph{arXiv preprint gr-qc/0411025} (2004) }.

\bibitem{cho2006asymptotic}
H.~Cho, \emph{Asymptotic quasinormal frequencies of different spin fields in
  spherically symmetric black holes}, {\emph{Physical Review D} {\bfseries 73}
  (2006) 024019}.

\bibitem{dolan2009expansion}
S.~R. Dolan and A.~C. Ottewill, \emph{On an expansion method for black hole
  quasinormal modes and regge poles}, {\emph{Classical and Quantum Gravity}
  {\bfseries 26} (2009) 225003}.

\bibitem{corda2015time}
C.~Corda, \emph{Time dependent schr{\"o}dinger equation for black hole
  evaporation: no information loss}, {\emph{Annals of Physics} {\bfseries 353}
  (2015) 71}.

\bibitem{jaramillo2021pseudospectrum}
J.~L. Jaramillo, R.~P. Macedo and L.~Al~Sheikh, \emph{Pseudospectrum and black
  hole quasinormal mode instability}, {\emph{Physical Review X} {\bfseries 11}
  (2021) 031003}.

\bibitem{Yang:2021civ}
R.-Q. Yang, R.-G. Cai and L.~Li, \emph{{Constraining the number of horizons
  with energy conditions}},
  \href{https://doi.org/10.1088/1361-6382/ac4118}{\emph{Class. Quant. Grav.}
  {\bfseries 39} (2022) 035005}.

\bibitem{Gwak:2022mze}
B.~Gwak, \emph{{Weak cosmic censorship conjecture in Myers-Perry black hole
  with separability}},
  \href{https://doi.org/10.1088/1475-7516/2022/10/077}{\emph{JCAP} {\bfseries
  10} (2022) 077}.

\bibitem{mamani2022revisiting}
L.~A. Mamani, A.~D. Masa, L.~T. Sanches and V.~T. Zanchin, \emph{Revisiting the
  quasinormal modes of the schwarzschild black hole: Numerical analysis},
  {\emph{The European Physical Journal C} {\bfseries 82} (2022) 897}.

\bibitem{kyutoku2023quasinormal}
K.~Kyutoku, H.~Motohashi and T.~Tanaka, \emph{Quasinormal modes of
  schwarzschild black holes on the real axis}, {\emph{Physical Review D}
  {\bfseries 107} (2023) 044012}.

\bibitem{barrabes1999metric}
C.~Barrabes, V.~Frolov and R.~Parentani, \emph{Metric fluctuation corrections
  to hawking radiation}, {\emph{Physical Review D} {\bfseries 59} (1999)
  124010}.

\bibitem{bellucci2010thermodynamic}
S.~Bellucci and B.~Tiwari, \emph{Thermodynamic geometry and hawking radiation},
  {\emph{Journal of High Energy Physics} {\bfseries 2010} (2010) 1}.

\bibitem{arias2012thermal}
E.~Arias, G.~Krein, G.~Menezes and N.~Svaiter, \emph{Thermal radiation from a
  fluctuating event horizon}, {\emph{International Journal of Modern Physics A}
  {\bfseries 27} (2012) 1250129}.

\bibitem{frolov2017quantum}
V.~P. Frolov and A.~Zelnikov, \emph{Quantum radiation from a sandwich black
  hole}, {\emph{Physical Review D} {\bfseries 95} (2017) 044042}.

\bibitem{barrabes2000stochastically}
C.~Barrabes, V.~Frolov and R.~Parentani, \emph{Stochastically fluctuating
  black-hole geometry, hawking radiation, and the trans-planckian problem},
  {\emph{Physical Review D} {\bfseries 62} (2000) 044020}.

\bibitem{hu2007metric}
B.~Hu and A.~Roura, \emph{Metric fluctuations of an evaporating black hole from
  backreaction of stress tensor fluctuations}, {\emph{Physical Review D}
  {\bfseries 76} (2007) 124018}.

\bibitem{hu2008stochastic}
B.~L. Hu and E.~Verdaguer, \emph{Stochastic gravity: Theory and applications},
  {\emph{Living Reviews in Relativity} {\bfseries 11} (2008) 1}.

\bibitem{emparan2008black}
R.~Emparan and H.~S. Reall, \emph{Black holes in higher dimensions},
  {\emph{Living Reviews in Relativity} {\bfseries 11} (2008) 1}.

\bibitem{horowitz2012black}
G.~T. Horowitz, \emph{Black holes in higher dimensions}. Cambridge University
  Press, 2012.

\bibitem{kanti2004black}
P.~Kanti, \emph{Black holes in theories with large extra dimensions: A review},
  {\emph{International journal of modern physics A} {\bfseries 19} (2004)
  4899}.

\bibitem{aharony2000large}
O.~Aharony, S.~S. Gubser, J.~Maldacena, H.~Ooguri and Y.~Oz, \emph{Large n
  field theories, string theory and gravity}, {\emph{Physics Reports}
  {\bfseries 323} (2000) 183}.

\bibitem{emparan2013large}
R.~Emparan, R.~Suzuki and K.~Tanabe, \emph{The large d limit of general
  relativity}, {\emph{Journal of High Energy Physics} {\bfseries 2013} (2013)
  1}.

\bibitem{emparan2015effective}
R.~Emparan, T.~Shiromizu, R.~Suzuki, K.~Tanabe and T.~Tanaka, \emph{Effective
  theory of black holes in the 1/d expansion}, {\emph{Journal of High Energy
  Physics} {\bfseries 2015} (2015) 1}.

\bibitem{emparan2020large}
R.~Emparan and C.~P. Herzog, \emph{Large d limit of einstein’s equations},
  {\emph{Reviews of Modern Physics} {\bfseries 92} (2020) 045005}.

\bibitem{licht2020black}
D.~Licht, R.~Luna and R.~Suzuki, \emph{Black ripples, flowers and dumbbells at
  large d}, {\emph{Journal of High Energy Physics} {\bfseries 2020} (2020) 1}.

\bibitem{suzuki2021squashed}
R.~Suzuki and S.~Tomizawa, \emph{Squashed black holes at large d},
  {\emph{Journal of High Energy Physics} {\bfseries 2021} (2021) 1}.

\bibitem{suzuki2022rotating}
R.~Suzuki and S.~Tomizawa, \emph{Rotating black holes at large d in
  einstein-gauss-bonnet theory}, {\emph{Physical Review D} {\bfseries 106}
  (2022) 024018}.

\bibitem{li2021ads}
P.-C. Li and C.-Y. Zhang, \emph{On ads black strings at large $ d$},
  {\emph{arXiv preprint arXiv:2112.11886} (2021) }.

\bibitem{licht2022lattice}
D.~Licht, R.~Luna and R.~Suzuki, \emph{Lattice black branes at large d},
  {\emph{Journal of High Energy Physics} {\bfseries 2022} (2022) 1}.

\bibitem{licht2022large}
D.~Licht, R.~Suzuki and B.~Way, \emph{The large d effective theory of black
  strings in ads}, {\emph{Journal of High Energy Physics} {\bfseries 2022}
  (2022) 1}.

\bibitem{suzuki2023phase}
R.~Suzuki and S.~Tomizawa, \emph{Phase and stability of black strings in
  einstein-gauss-bonnet theory at large d}, {\emph{Journal of High Energy
  Physics} {\bfseries 2023} (2023) 1}.

\bibitem{andrade2020entropy}
T.~Andrade, R.~Emparan, A.~Jansen, D.~Licht, R.~Luna and R.~Suzuki,
  \emph{Entropy production and entropic attractors in black hole fusion and
  fission}, {\emph{Journal of High Energy Physics} {\bfseries 2020} (2020) 1}.

\bibitem{mandlik2021black}
M.~Mandlik, \emph{Black rings in large d membrane paradigm at the first order},
  {\emph{Journal of High Energy Physics} {\bfseries 2021} (2021) 1}.

\bibitem{mandlik2022sitter}
M.~Mandlik, \emph{De sitter static black ring in large d membrane paradigm at
  the second order}, {\emph{Journal of High Energy Physics} {\bfseries 2022}
  (2022) 1}.

\bibitem{kachru2023holographic}
S.~Kachru and M.~Shyani, \emph{Holographic non-fermi liquids at large d},
  {\emph{Journal of High Energy Physics} {\bfseries 2023} (2023) 1}.

\bibitem{keeler2022hidden}
C.~Keeler, V.~Martin and A.~Priya, \emph{Hidden conformal symmetries from
  killing towers with an application to large-d/cft}, {\emph{SciPost Physics}
  {\bfseries 12} (2022) 170}.

\bibitem{giataganas2022holographic}
D.~Giataganas, N.~Pappas and N.~Toumbas, \emph{Holographic observables at large
  d}, {\emph{Physical Review D} {\bfseries 105} (2022) 026016}.

\bibitem{kirezli2022classification}
P.~Kirezli, \emph{Classification of robinson-trautman and kundt geometries with
  large d limit}, {\emph{Journal of High Energy Physics} {\bfseries 2022}
  (2022) 1}.

\bibitem{sybesma2023zoo}
W.~Sybesma, \emph{A zoo of deformed jackiw-teitelboim models near large
  dimensional black holes}, {\emph{Journal of High Energy Physics} {\bfseries
  2023} (2023) 1}.

\bibitem{luna2023holographic}
R.~Luna and M.~Sanchez-Garitaonandia, \emph{Holographic collisions in large d
  effective theory}, {\emph{Journal of High Energy Physics} {\bfseries 2023}
  (2023) 1}.

\bibitem{Barrabes:2013zva}
C.~Barrab\`es and P.~A. Hogan, \emph{{Advanced general relativity : gravity
  waves spinning particles and black holes}}. Oxford U. Pr., Oxford, 2013,
  \href{https://doi.org/10.1093/acprof:oso/9780199680696.001.0001}{10.1093/acprof:oso/9780199680696.001.0001}.

\bibitem{tangherlini1963schwarzschild}
F.~R. Tangherlini, \emph{Schwarzschild field in n dimensions and the
  dimensionality of space problem}, {\emph{Il Nuovo Cimento (1955-1965)}
  {\bfseries 27} (1963) 636}.

\bibitem{iyer1989vaidya}
B.~Iyer and C.~Vishveshwara, \emph{The vaidya solution in higher dimensions},
  {\emph{Pramana} {\bfseries 32} (1989) 749}.

\bibitem{zhao2017spherical}
L.~Zhao, \emph{Spherical and spheroidal harmonics: Examples and computations},
  Ph.D. thesis, The Ohio State University, 2017.

\bibitem{erdelyi1953higher}
A.~Erdelyi, W.~Magnus, F.~Oberhettinger and F.~Tricomi, \emph{Higher
  transcendental functions mcgraw-hill}, {\emph{New York} {\bfseries 2} (1953)
  }.

\bibitem{berti2004gravitational}
E.~Berti, M.~Cavaglia and L.~Gualtieri, \emph{Gravitational energy loss in high
  energy particle collisions: Ultrarelativistic plunge into a multidimensional
  black hole}, {\emph{Physical Review D} {\bfseries 69} (2004) 124011}.

\end{thebibliography}\endgroup
\end{document}